\documentclass{JINST}
\usepackage{graphicx}
\usepackage{txfonts}
\usepackage{natbib}

\newcommand{\OMISSIS}[1]{\begin{center}{$\mathbf{[\dots \: \mathrm{OMISSIS} \: \dots]}$}\end{center}}

\def\PTypeZ{PType~0}
\def\PTypeI{PType~1}
\def\PTypeII{PType~2}
\def\PTypeIII{PType~3}
\def\PTypeIV{PType~4}
\def\PTypeV{PType~5}
\def\PTypeVI{PType~6}

\def\Type{{\tt Type}}
\def\SubType{{\tt Subtype}}

\def\scos{{\tt{SCOS}}~2000}
\def\TMH{{\tt TMH}}

\def\LONE{{\tt L1}}

\def\ismp{i_{\mathrm{smp}}}

\def\DAEGain{G_{\mathrm{DAE}}}
\def\DAEZero{Z_{\mathrm{DAE}}}
\def\DAEOffset{O_{\mathrm{DAE}}}
\def\Vadc{V_{\mathrm{adc}}}

\def\Vin{V_{\mathrm{in}}}
\def\Vout{V_{\mathrm{out}}}

\def\qADC{q_{\mathrm{ADC}}}

\def\DSA{\mathrm{DSA}}
\def\BTI{\mathrm{BTI}}
\def\DGI{\mathrm{DGI}}
\def\DOI{\mathrm{DOI}}

\def\Vin{V_{\mathrm{in}}}

\def\V{\mathrm{V}}

\def\OBT{t^{\mathrm{obt}}}

\def\REBAnaver{\mathrm{N_{aver}}}
\def\REBAq{\mathrm{SECOND\_QUANT}}
\def\REBAo{\mathrm{OFFSET\_ADJUST}}
\def\REBAgo{\mathrm{GMF1}}
\def\REBAgt{\mathrm{GMF2}}

 \def\Planck{{\sc Planck}}
 \def\sky{sky}
 \def\load{reference--load}

 \def\onboard{on--board}

 \newcommand{\freqsampling}{f_{\mathrm{sampling}}}     

 \newcommand{\Naver}{N_{\mathrm{aver}}}     

 \newcommand{\Pone}{P_{1}} 
 \newcommand{\Ptwo}{P_{2}} 

 \newcommand{\round}{\mathrm{round}}

 \newcommand{\trunc}{{\mathrm{trunc}}}

\def\FIGPTYPES{
\begin{figure*}[hbt!]
 \centering
 \includegraphics[width=0.75\textwidth]{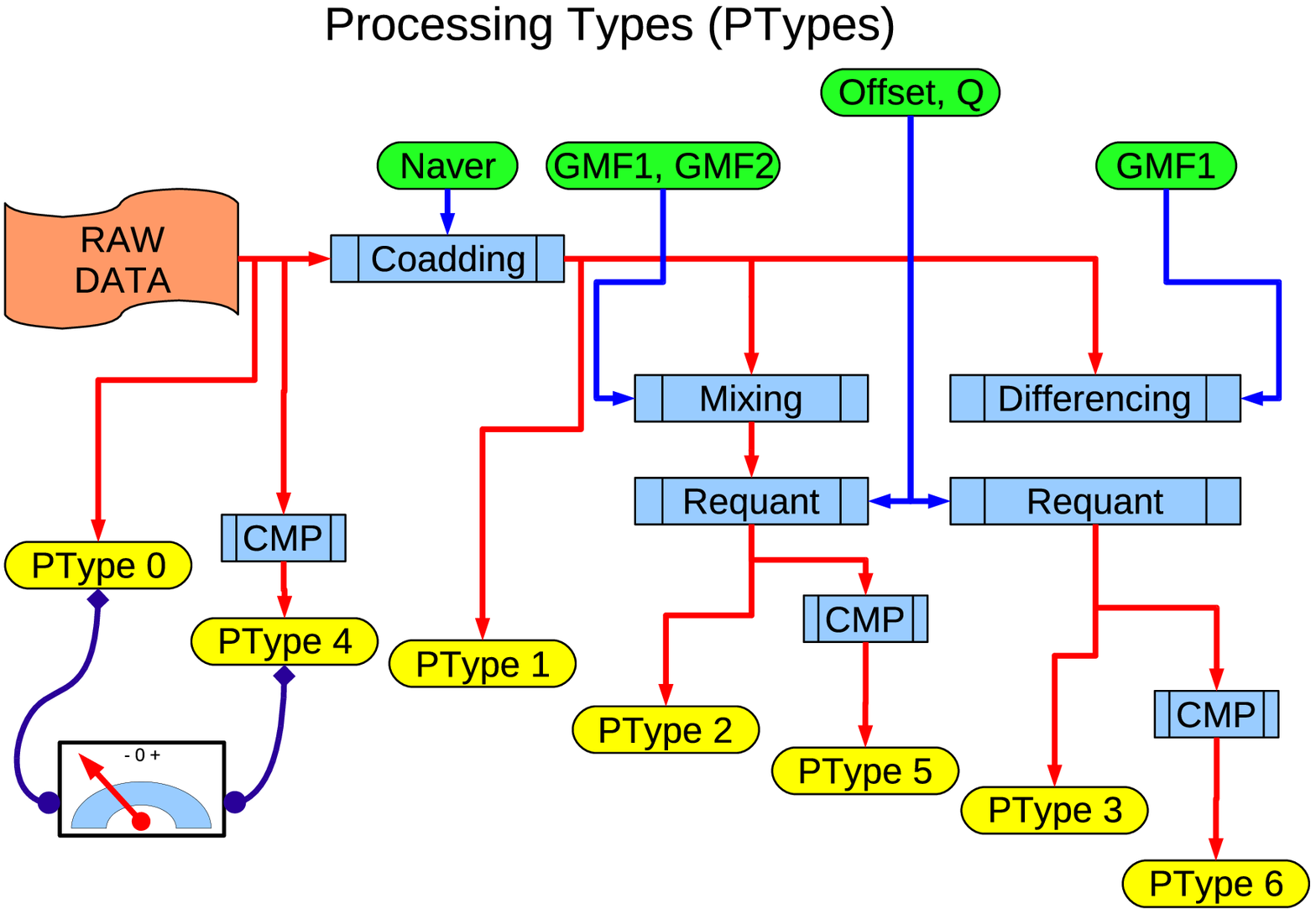}
\caption{
Schematic representation of the scientific \onboard\ processing, processing
parameters and processing types for the LFI. The diagram shows the sequence of
operations leading to each processing type: coadding, mixing, requantization
(Requant) and compression (CMP).
\label{fig:ptypes}
}
\end{figure*}
} 

\def\FIGPTYPESPROCESSING{
\begin{figure*}
 \centering
 \includegraphics[angle=-90,width=0.7\textwidth]{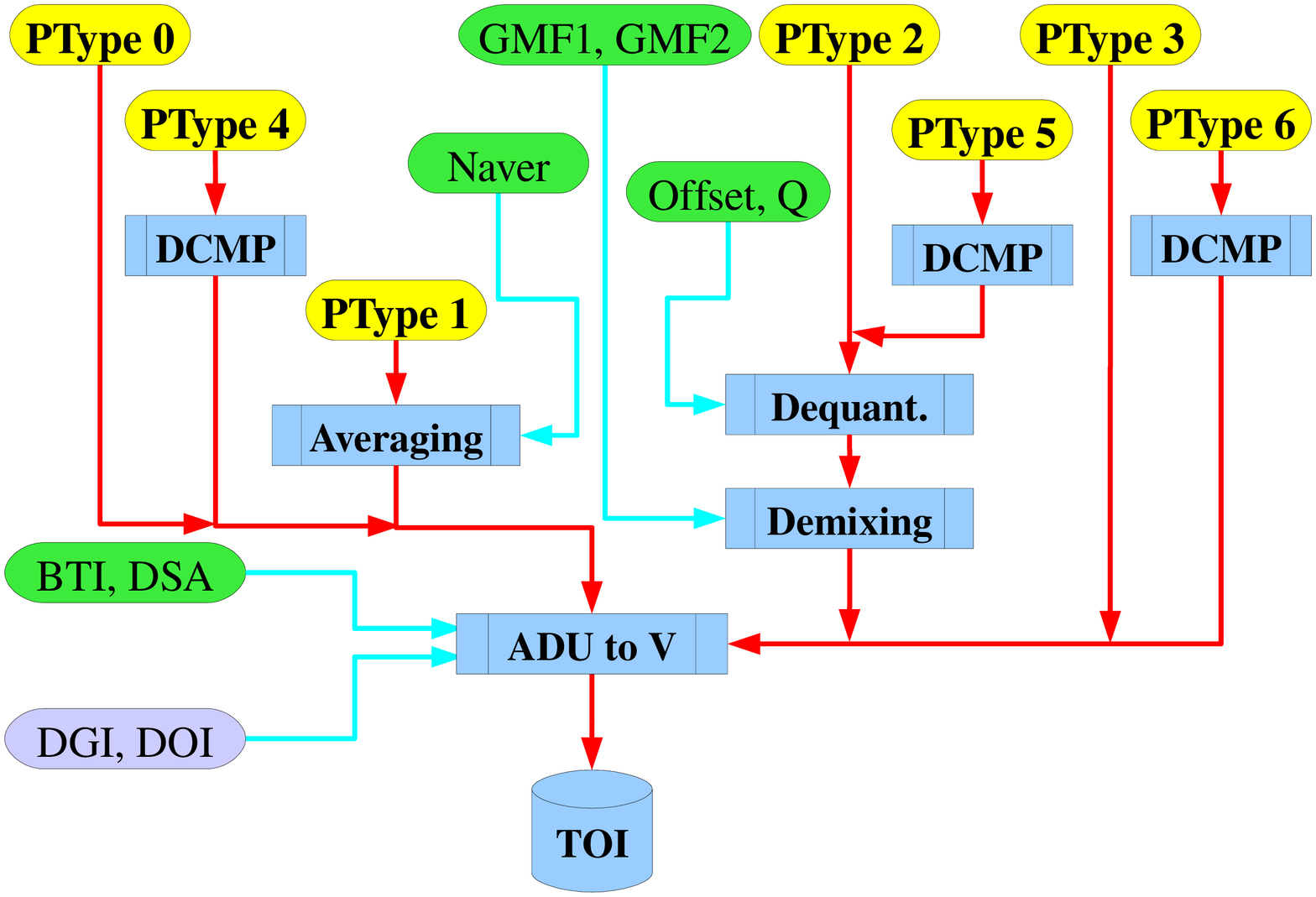}
\vspace{-0.05\textwidth}
\caption{
 Schematic representation of the \LONE\ processing for the various processing parameters and processing types for the LFI. 
 $\REBAnaver$, $\REBAgo$, $\REBAgt$, $\REBAo$ (abbreviated with Offset), $\REBAq$ (abbreviated with Q), BTI and DSA are
recovered from the tertiary header of each packet, DGI and DOI are recovered from the HK telemetry.
\label{fig:ptypes:processing}
}
\end{figure*}
} 

\def\FIGPACKETS{
\begin{figure}[htb!]
 \centering
 \includegraphics[angle=-90,width=0.45\textwidth]{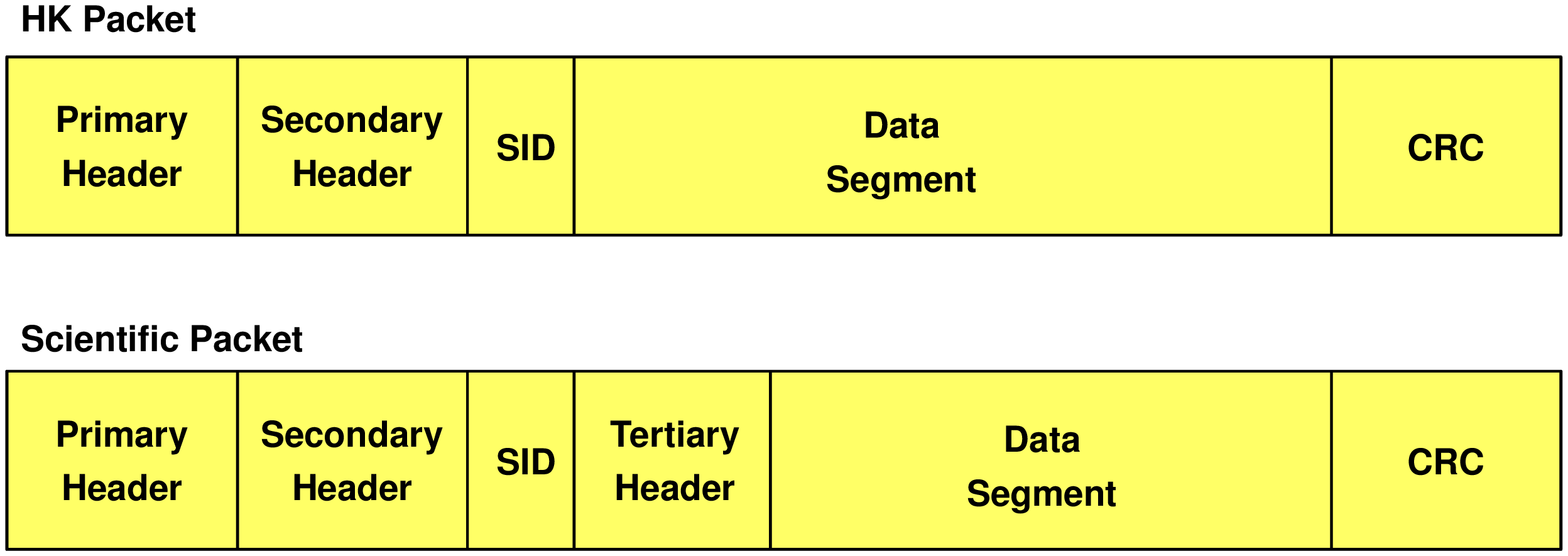}
\vspace{-0.15\textwidth}
\caption{
General structure of scientific and HK telemetry packets.
\label{fig:packets}
}
\end{figure}
} 

\def\FIGDATASEGMENT{
\begin{figure}[htb!]
 \centering
 \includegraphics[angle=-90,width=0.5\textwidth]{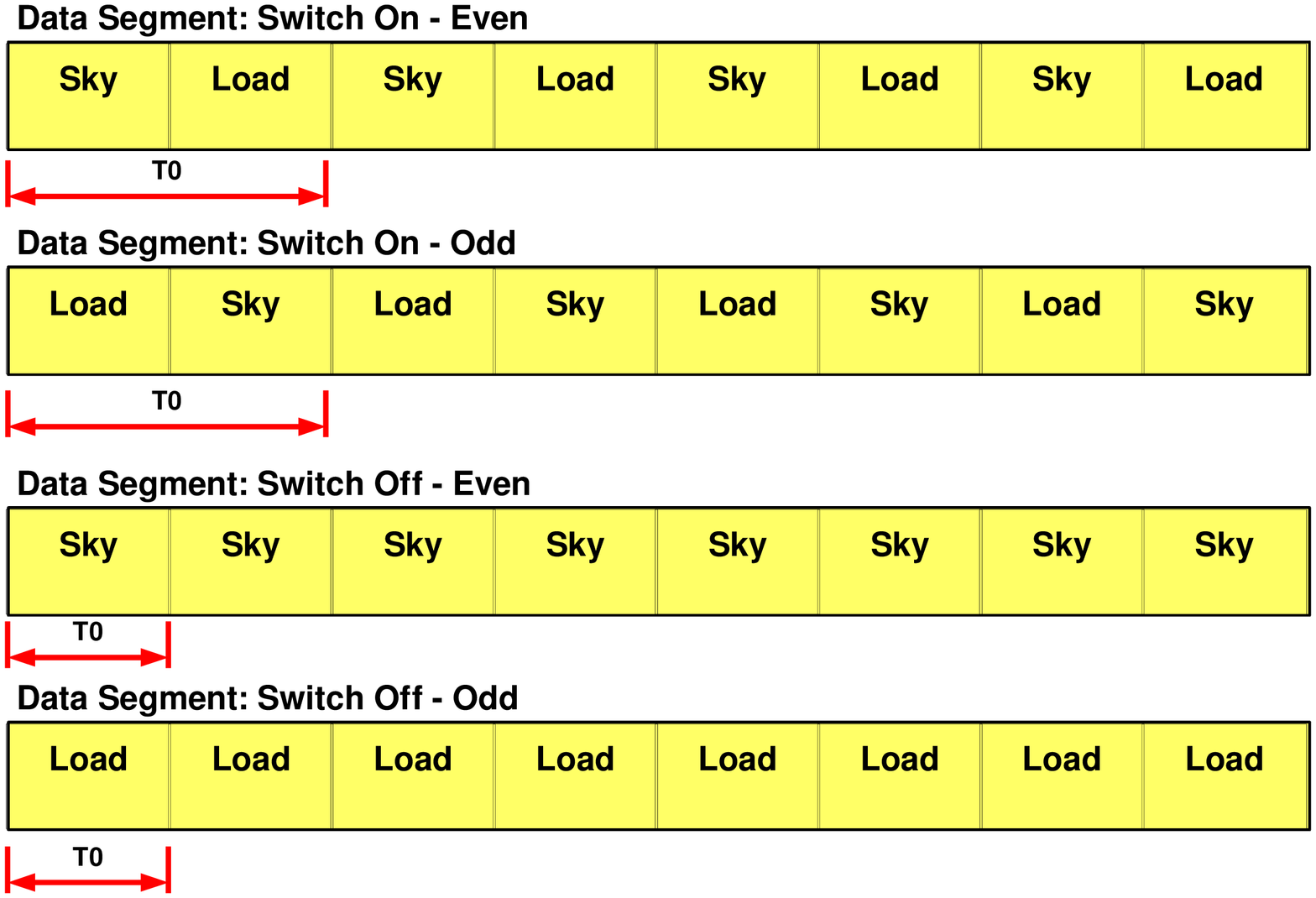}
\vspace{-0.05\textwidth}
\caption{
Content of the data segment for scientific packets of \PTypeZ, 1, 2, 4, 5 after decompression and demixing (if needed).
From top to bottom, packet with phase switch on, at even phase and odd phase,
and with phase switch off, with even and odd phases. The double arrow with $T0$ denotes the samples with $\OBT$\ given by the packet timestamp.
For PType $\ne 0$ and phase switch on, consecutive couples of samples have the same timestamp.
When the phase switch is off, each sample has its hown $\OBT$.
\label{fig:data:segment}
}
\end{figure}
} 

\def\FIGLEVEL1ARCH{
\begin{figure*}
 \centering
 \includegraphics[width=0.8\textwidth]{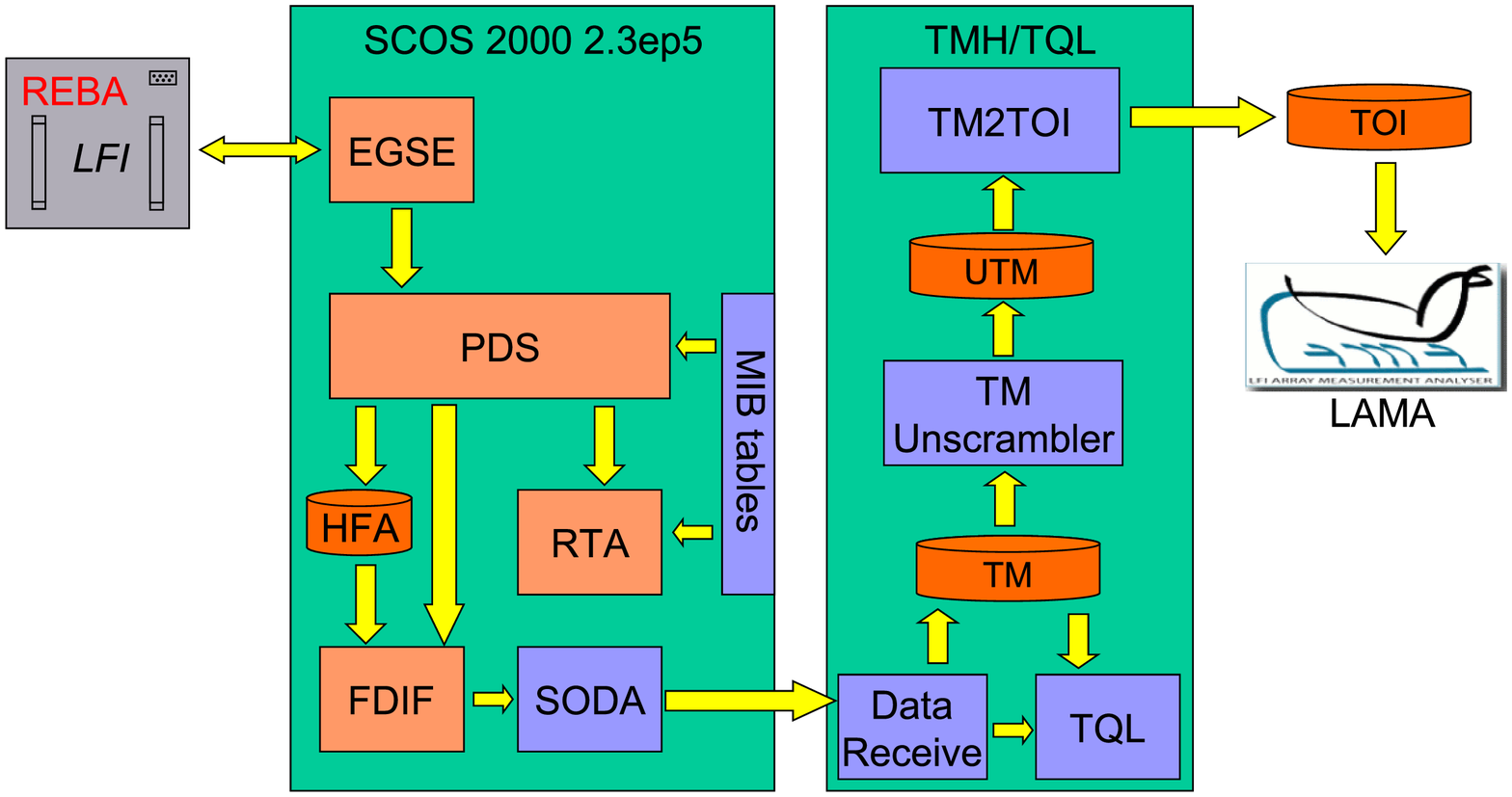}
\caption{
Level 1 architecture and data flow, showing the main components of each
subsystem. The EGSE subsystem of SCOS receives the
telemetry data from the REBA. The Packet Distribution System (PDS)
forwards the telemetry packets to the Real-Time Assessment (RTA) displays,
the History File Archive (HFA) and to the Flight Dynamic Interface (FDIF). 
Packets are properly parsed using the information stored in the Mission
Information Base (MIB). The SODA task receives the real-time telemetry through
the FDIF interface and forwards it to the TMH/TQL system. The off-line
analysis of the TOI generated by the TMH, during the LFI calibration campaign,
was performed using the LAMA software \citep{Tom09}. 
\label{fig:level1arch}
}
\end{figure*}
} 

\def\FIGTQL{
\begin{figure*}
 \centering
 \includegraphics[width=0.8\textwidth]{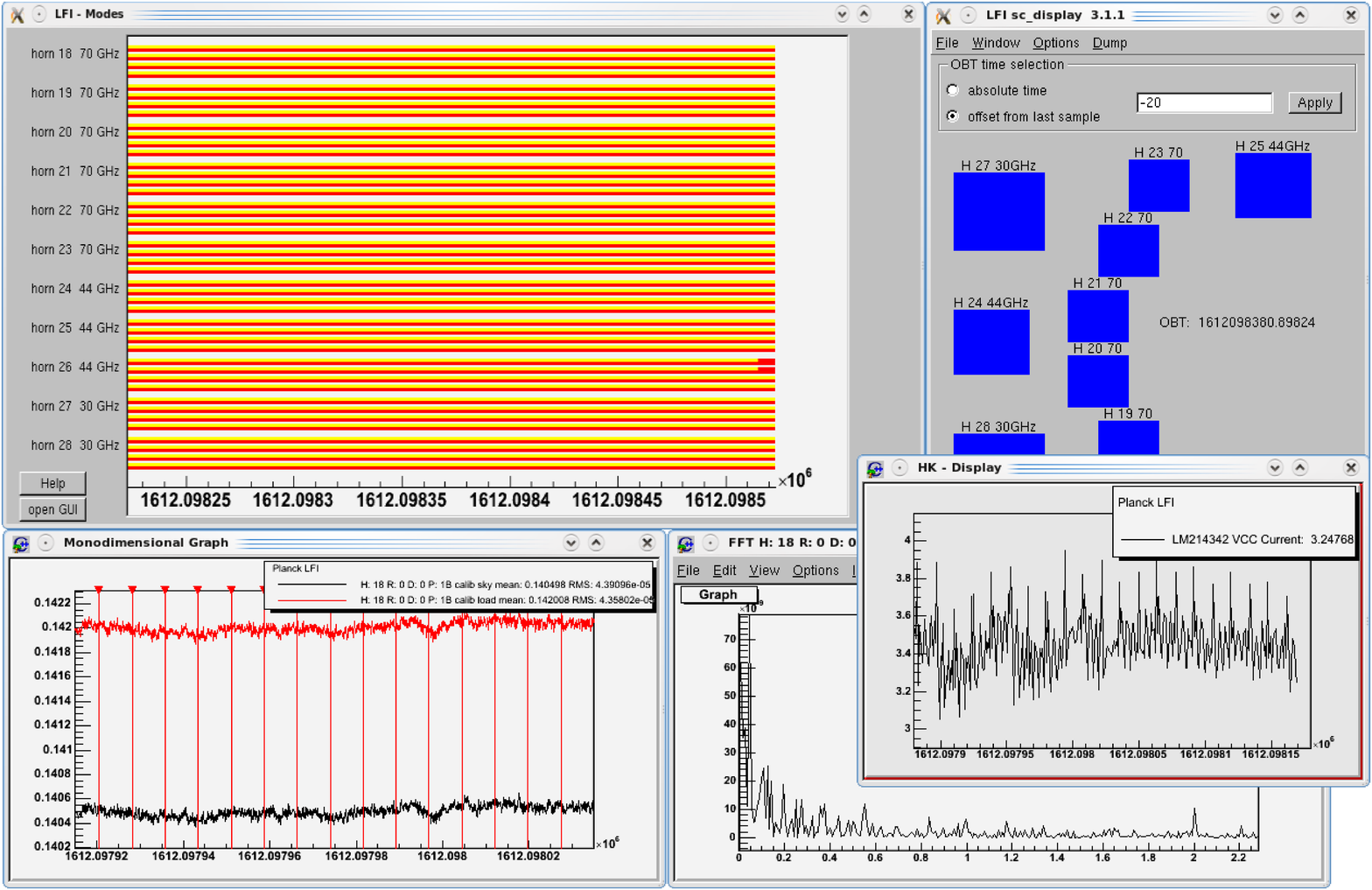}
\caption{
The TQL displays. The LFI modes display (top left) shows the processing type for
all detectors of each horn. The status display window (top right) shows the
status of all detectors of all horns, reporting if there is no data for a given
detector or if normal scientific data or calibration data has been received at a
given time. The mono-dimensional graph (bottom left) displays the load and sky
data of one or more detectors as a function of time. On the bottom right two
displays are shown: the fast Fourier transform of the sky signal for a single
detector of feed-horn 18 (in background) and the HK display plotting a single LFI
parameter as a function of time (in foreground).
\label{fig:tql}
}
\end{figure*}
} 

\def\FIGTMU{
\begin{figure*}
 \centering
 \includegraphics[width=0.8\textwidth]{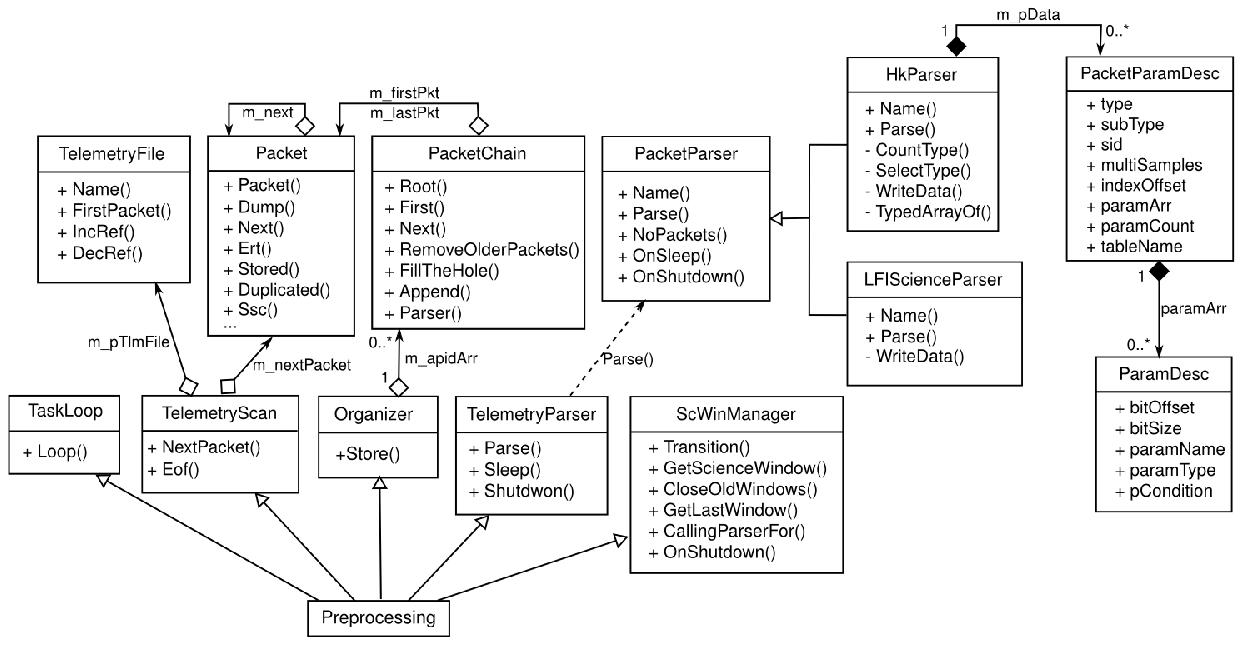}
\caption{
The TMU class diagram (in UML).
\label{fig:tmu}
}
\end{figure*}
} 

\def\instOATS{$^{a}$}
\def\instISDC{$^{b}$}
\def\instIASFBO{$^{c}$}
\def\instSISSA{$^{d}$}
\def\instUNIMI{$^{e}$}
\def\instUNITS{$^{f}$}
\def\instJBO{$^{g}$}
\def\instUSA{$^{h}$}
\def\instESAC{$^{i}$}
\def\instCNR{$^{j}$}

\title{Level 1 on-ground telemetry handling in {\sc Planck} LFI
\footnote { Submitted to JINST: 24 June 2009, Accepted: 28 August
2009, Published 29 December 2009.
 Reference : 2009 JINST 4 T12019
 DOI: 10.1088/1748-0221/4/12/T12019
  }
}

\author{
A.~Zacchei{\instOATS} \thanks{Corresponding Author,
e--mail:zacchei@oats.inaf.it} , M.~Frailis{\instOATS},
M.~Maris{\instOATS}, N.~Morisset{\instISDC}, R.~Rohlfs{\instISDC},
M.~Meharga{\instISDC}, P.~Binko{\instISDC},
M.~T\"urler{\instISDC}, S.~Galeotta{\instOATS},
F.~Gasparo{\instOATS}, E.~Franceschi{\instIASFBO},
R.~C.~Butler{\instIASFBO}, F.~Cuttaia{\instIASFBO},
O.~D'Arcangelo{\instCNR}, S.~Fogliani{\instOATS},
A.~Gregorio{\instUNITS}, R.~Leonardi{\instUSA},
S.R.~Lowe{\instJBO}, D.~Maino{\instUNIMI}, G.~Maggio{\instOATS},
M.~Malaspina{\instIASFBO}, N.~Mandolesi{\instIASFBO},
P.~Manzato{\instOATS}, P.~Meinhold{\instUSA},
L.~Mendes{\instESAC}, A.~Mennella{\instUNIMI},
G.~Morgante{\instIASFBO}, F.~Pasian{\instOATS},
F.~Perrotta{\instSISSA}, M.~Sandri{\instIASFBO},
L.~Stringhetti{\instIASFBO}, L.~Terenzi{\instIASFBO},
M.~Tomasi{\instUNIMI}, A.~Zonca{\instUNIMI} \\
\llap{\instOATS} Osservatorio Astronomico di Trieste, INAF,\\
 Via G.B.~Tiepolo 11, I-34131, Trieste, Italy\\
\llap{\instISDC} University of Geneva, \\
 ch. d'Ecogia 16, 1290 Versoix, Switzerland\\
\llap{\instIASFBO} IASF - Sezione di Bologna, INAF,\\
 Via P.~Gobetti, 101, I-40129 Bologna, Italy\\
\llap{\instSISSA} SISSA - ISAS,\\
 via Beirut 2-4, I-34151 Trieste, Italy\\
\llap{\instUNIMI} Dipartimento di Fisica, Universit\'a degli studi di Milano,\\
 Via G.~Celoria 16, I-20133 Milano, Italy\\
\llap{\instUNITS} Dipartimento di Fisica, Universit\`a degli Studi di Trieste,\\
 Via Valerio 2, I-34127 Trieste, Italy\\
\llap{\instJBO} Jodrell Bank Centre for Astrophysics, The University of Manchester,\\
 Manchester, M60 1QD, UK \\
\llap{\instUSA} Department of Physics, University of California,\\
 Santa Barbara, CA 93106-9530, USA\\
\llap{\instESAC} ESA - ESAC\\
 Camino bajo del Castillo, s/n, Villanueva de la Ca\~{n}ada 28692 Madrid\\
\llap{\instCNR} Istituto di Fisica del Plasma, CNR,\\
 Via Cozzi 53, Milano, Italy \\
  }

\begin{abstract}
  {The \Planck\ Low Frequency Instrument (LFI) will observe the Cosmic Microwave
    Background (CMB) by covering the frequency range 30-70 GHz in three
    bands. The primary instrument data source are the temperature samples
    acquired by the 22 radiometers mounted on the Planck focal plane. Such
    samples represent the scientific data of LFI. In addition, the LFI
    instrument generates the so called housekeeping data by sampling regularly
    the on-board sensors and registers. The housekeeping data provides
    information on the overall health status of the instrument and on the
    scientific data quality. The scientific and housekeeping data are collected
    on-board into telemetry packets compliant with the ESA Packet Telemetry
    standards. They represent the primary input to the first processing level of
    the LFI Data Processing Centre. In this work we show the software systems
    which build the LFI Level 1. A real-time assessment system, based on the ESA
    SCOS 2000 generic mission control system, has the main purpose of monitoring
    the housekeeping parameters of LFI and detect possible anomalies. A
    telemetry handler system processes the housekeeping and scientific telemetry
    of LFI, generating timelines for each acquisition chain and each
    housekeeping parameter. Such timelines represent the main input to the
    subsequent processing levels of the LFI DPC. A telemetry quick-look system
    allows the real-time visualization of the LFI scientific and housekeeping
    data, by also calculating quick statistical functions and fast Fourier
    transforms. The LFI Level 1 has been designed to support all the mission
    phases, from the instrument ground tests and calibration to the flight
    operations, and developed according to the ESA engineering standards.}
\end{abstract}


\begin{document}



\def\sep{--}


 \maketitle

 \section{Introduction}\label{sec:introduction}

The Planck Science Ground Segment comprises the LFI and HFI Data
Processing Centers (DPCs), the Mission Operations Control Centre
(MOC) and the Planck Science Office (PSO). The LFI DPC is
responsible of the data reduction and scientific processing of the
LFI instrument, in order to extract the maximum amount of
scientific information and provide the data quality necessary to
achieve the objectives of the Planck mission \citep{Tau04}.

The LFI DPC activities have been decomposed into 4 levels, each one
characterized by a subset of the data products generated during the Planck
mission. In particular, the main input of the Level 1 is the raw telemetry of
the instrument, compliant with the format specified by the ESA Packet Telemetry
Standard \citep{ESA88} and the Packet Utilization Standard
\citep{ESA02}. Additional auxiliary input data is necessary to associate the
telemetry data to a precise pointing of Planck and to correlate the on-board
time (OBT) with the Universal Time (UTC) during flight operations. The Level 1
produces as output Time Ordered Information (TOI), i.e. time series, of each
housekeeping and scientific parameter of LFI, together with additional
diagnostic information necessary to monitor the health of the instrument and the
data quality. The TOI are the primary input of the Level 2 and are also part of
the final products of the LFI DPC versus the scientific community (after
calibration and removal of systematic effects). They will also be used to create
maps of the main astrophysical components.

This paper describes the software systems which are part of the Level 1 of
Planck/LFI. A \textbf{Real-Time Assessment system} (\textbf{RTA}), based on the
SCOS 2000 generic mission control system of ESA, provides monitoring tools to
perform a routine analysis of the Spacecraft (S/C) and Payload (P/L)
Housekeeping (HK) telemetry.  A \textbf{Telemetry Handling system}
(\textbf{TMH}) processes the raw telemetry packets by extracting the HK
parameters and by uncompressing and decoding the LFI scientific samples (SCI);
it produces homogeneous timelines for each HK and SCI parameter. A timeline
contains for each sample, the OBT time, its raw and calibrated value together
with additional flags on the data quality. A \textbf{Telemetry Quick-Look
  system} (\textbf{TQL}) provides interactive quick-look analysis tools,
specific for the LFI instrument, to check both the SCI and HK telemetry and
verify the normal behavior of the instrument.

The Level 1 software has been designed and developed in order to
support all phases of the Planck/LFI mission, from the instrument
ground tests, tuning and calibration to the integration tests and
the flight operations. Priority has been given to the development
of the core functionalities, common to all mission phases, and to
maximizing code reuse. In this paper, we describe the software
versions in use during the ground test campaigns (the so called
Flight Models). The verification and validation of the Level 1
software, performed according to the ESA software engineering
standard, has been described in \citep{Fra09}.

 \section{The LFI on-board data processing and telemetry structure}\label{sec:LFI:acqusition}

 \subsection{The on-board processing}

In order to fully understand the operations performed by the
Level 1 software, and in particular by the TMH/TQL systems, on the LFI scientific
telemetry, we first provide an overview of the LFI acquisition chain and
on-board processing.

The heart of the LFI instrument \citep{Ber09} is an array of 22 radiometers
fed by 11 corrugated feedhorns arranged in a circular
pattern in the Planck focal plane. The receivers operate at three well separated
bands: 30, 44 and 70 GHz. For each feedhorn, an ortho-mode transducer separates
the two orthogonal polarization components. Each radiometer is a differential
pseudo-correlation receiver \citep{Men03} where the signals from the sky and
from a reference load held at the constant temperature of 4~K are combined by a
hybrid coupler, amplified in two independent amplifier chains, passed through a
phase switch and separated out by another hybrid. The phase switch adds 0/180
degrees of phase lag to the signals and it is switched at 4~kHz synchronously
in one of the two amplifier chains.
%
%

Hence, each radiometer provides two analog outputs, one for each amplifier
chain. In a nominal configuration, each output yields a sequence of alternating
$T_{load}$,$T_{sky}$ signals at the frequency of the phase switch. By changing the
phase switches configuration, the output can be a sequence of either $T_{sky}$ or
$T_{load}$ signals.

The conversion from analog to digital form of each radiometer output is performed by a
14~bits Analog-to-Digital Converter (ADC) in the Data Acquisition Electronics
unit (DAE). The DAE transforms the signal in the range $[-2.5 V, +2.5 V]$: first
it applies a tunable \textbf{offset}, $\DAEOffset$, then it amplifies the signal
with a tunable \textbf{gain}, $\DAEGain$, in order to make full use of the
resolution of the ADC, and finally the signal is integrated. To eliminate
phase switch raise transients, the integration takes into account a {\bf
  blanking time}, i.e. a blind time in the integrator where data are not
considered. The default value of the blanking time is 7.5 $\mu$s. Both the
$\DAEOffset$, the $\DAEGain$ and the blanking time are parameters set through
the LFI on-board software. The equation applied to transform a given input
signal $\Vin$ into an output $\Vout$ is

\begin{equation}
 \Vout = \DAEGain (\Vin + \DAEOffset) + \DAEZero
\end{equation}

 \noindent
with
$\DAEGain$ = 1, 2, 3, 4, 6, 8, 12, 16, 24, 48,
$\DAEOffset$ is one of 255 possible offset steps from +0 up to +2.5 V and
where $\DAEZero$ is a small offset introduced by the DAE when applying a
selected gain. The values of $\DAEGain$ and $\DAEOffset$ are set by
sending, through specific telecommands, the DAE Gain Index (DGI) and the DAE
Offset Index (DOI) associated to the desired values. A calibration table is
needed for each offset step in order to best estimate how many volts have been
removed.

The ADC quantizes the $\Vout$ uniformly in the range $-2.5 \V \le \Vadc \le +2.5
\V$, so that the quantization step is $\qADC=0.30518$~mV. The quantization
formula is

  \begin{equation}
   P = \round \left[\frac{\Vout}{\qADC} \right],
  \end{equation}

 \noindent
and the output is stored as an unsigned integer of 16 bits.

The digitized scientific data is then processed by the Radiometer Electronics
Box Assembly (REBA) which runs the LFI on-board software. The REBA is also in
charge of processing the HK data (e.g. digital conversion of temperatures and
voltages), keeping the OBT register synchronized with the S/C on-board
clock, the generation of the LFI HK and SCI telemetry packets, the
interpretation and execution of the telecommands sent from ground to the
Spacecraft and in general the management of the entire instrument.

\FIGPTYPES

According to the ESA Standards, a source packet must include, in addition to the
source data, a minimum of information needed by the ground system for the
acquisition, storage and distribution of the source data to the end user. The
source data is divided into several telemetry packets if it exceeds a prescribed
maximum length.

Hence, the REBA processes data from each on-board {\em data source} in form of
time series which are sampled at discrete times and split into packets to be
sent to Earth. Since, in the Planck satellite, the telemetry packets of an
entire operational day are stored into an on-board mass memory and then
transmitted to the ground station during the Daily Tele-Communication Period
(DTCP), the telemetry budget of the LFI instrument must not exceed the limit of
53.5 Kbps. To satisfy this limit, the REBA implements 7 acquisition modes
(processing types) which reduce the scientific data rate by applying a number of
processing steps. Fig.~\ref{fig:ptypes} illustrates the main steps of the
\onboard\ processing and the corresponding processing types (PTypes):

\begin{itemize}
\item {\bf \PTypeZ}: in this mode the REBA just packs the raw data of the
  selected channel without any processing.
\item {\bf \PTypeI}: consecutive \sky\ or \load\ samples are coadded and stored
  as unsigned integers of 32 bits. The number of consecutive samples to be
  coadded is specified by the $\REBAnaver$ parameter.
\item {\bf \PTypeII}: in this mode, two main processing steps are applied. First,
  pairs of averaged \sky\ and \load\ samples, respectively $\overline{S}_{sky}$ and
  $\overline{S}_{load}$, are mixed by applying two gain modulation factors, $\REBAgo$ and
  $\REBAgt$:
\begin{eqnarray}
   \Pone & = & \overline{S}_{sky} - \REBAgo \cdot \overline{S}_{load} \\
   \Ptwo & = & \overline{S}_{sky} - \REBAgt \cdot \overline{S}_{load}
\end{eqnarray}
The operations are performed as floating point operations. Then the values
obtained are requantized converting them into 16-bit signed integers:

\begin{equation}
  Q_i = \round\left[ \REBAq \left( P_i + \REBAo \right) \right]
\end{equation}
\item {\bf \PTypeIII}: with respect to \PTypeII, in this mode only a single gain
  modulation factor is used, $\REBAgo$, obtaining:
\begin{eqnarray}
   P & = & \overline{S}_{sky} - \REBAgo \cdot \overline{S}_{load}
\end{eqnarray}
and analogously to \PTypeII, the value is  requantized obtaining 16-bit
signed integer.
\item With the processing types {\bf \PTypeIV}, {\bf \PTypeV}, {\bf \PTypeVI} the
  REBA performs a loss-less adaptive arithmetic compression of the data obtained
  respectively with the processing types \PTypeZ, \PTypeII\ and \PTypeIII. The
  compressor takes couples of 16 bit numbers and stores them in the output
  string up to the complete filling of the data segment for the packet in
  process.
\end{itemize}

The set of {\em REBA parameters} -- $\REBAnaver$, $\REBAgo$,
$\REBAgt$, $\REBAq$ and $\REBAo$ --\ can be selected for each of
the 44 LFI channels independently by sending dedicated
telecommands. The calibration of the REBA parameters has been
detailed in \citep{Mar09}.The REBA can acquire data from a channel
in two modes at the same time. This capability is used to verify
the effect of a certain processing type on the data quality. So,
in nominal conditions, the LFI instrument will use the \PTypeV\
for all its 44 detectors and every 15 minutes a single detector,
in turn, will be processed also with \PTypeI\ in order to
periodically calibrate the gain modulation factors and the second
quantization. The other processing types are mainly used for
diagnostic, testing or contingency purposes.

The HK data processing is simpler since, basically, all the HK sources are
polled at fixed times and analog sources are converted into digital form through
an 8-bits linear ADC. The HK parameters are then grouped by: sub-system,
information status (essential, nominal), purpose (monitoring, diagnostic) and
sampling rate. Each group is tagged with a single OBT time and stored in the
same telemetry packet.

In order to properly transform a data segment within a telemetry packet into a
time series, each SCI and HK packet contains a time field associated to the
first sample of the data segment. A clock internal to the REBA generates the
sampling periods and the instrument OBT. Unavoidable drifts
in the \onboard\ clock lead to a loss of precise synchronization between the
OBT and the universal time. For this reason the OBT is converted at ground to
UTC (Universal Time Correlated).

\subsubsection{Telemetry Packet Structure}
Packets generated by the REBA follow the ESA Packet Telemetry Standard and
Packet Telecommand Standard, the CCSDS Packet Telemetry recommendations
\citep{CCS95} and the ESA Packet Utilization Standard
(PUS). Fig.~\ref{fig:packets} shows the general structure of the HK and SCI
telemetry packets while Fig.~\ref{fig:data:segment} shows the general structure
of a scientific data segment.

Packets are made of two headers of fixed length, a data segment of variable
length and a 16~bits CRC field used to check the consistency of the packet. The
specification of the Cyclic Redunndacy Code is provided in the ESA PUS, together
with a reference implementation in C language.

All the packets have a primary header, carrying information about the type of
packet (telecommand or telemetry) as well as a unique identifier of the on-board
application process that has generated it (APID), a sequence counter and the
total length of the packet. For telemetry packets, a secondary header of fixed
length carries information about the type of service generating the packet
(e.g. TC Verification, HK data reporting, Event reporting) and the on-board
reference time of the packet.

\FIGPACKETS

\FIGDATASEGMENT

The sequence counter of the packet contains a sequential count (modulo 16384) of
each packet by each source application process on the spacecraft. It allows the
ground segment to detect possible gaps in the telemetry flow. The OBT is stored
as a 48-bit field in CUC format, as specified in the PUS.

The aforementioned services are group of functionalities to be implemented
on-board the satellite. Standard services are defined by the ESA PUS and are
uniquely identified by a couplet of numbers (\Type,~\SubType). A service is
invoked by a service request (telecommand source packet) which will result in
the generation of zero or more service reports (telemetry source packet). Since
a single service can generate several types of telemetry packets, with different
structures, additional fields within the packet can be necessary to uniquely
identify and properly parse a telemetry packet. Hence, in Planck most of the TM
packets have a Structure ID field (SID). The set of fields (APID,
\Type,~\SubType, SID) within a packet is used by the ground software to identify
the data segment structure and extract the HK or the SCI samples.

Since the ESA PUS addresses only the utilization of telecommand packets and
telemetry source packets for the purpose of remote monitoring and control of
subsystems and payloads, an additional custom set of services has been defined
by the PLANCK mission in order to handle the scientific telemetry.

\subsubsection{Mapping SCI data into packets}
After the SID, SCI telemetry packets contain a tertiary header of fixed length
that includes the identifier of the detector that has generated the data, the
phase switch status, the processing type applied to the data, the REBA
parameters used for the processing, and the total number of samples stored in
the packet \citep{Gre08}.

For the processing types \PTypeZ, 1 and 4 the switching status defines whether
the packet contains a sequence of interleaved \sky\ and \load\ samples, or a
sequence of only \sky\ or only \load\ data.  For PTypes 2 or 5 the switching
status specifies the kind of sequence of samples produced by the demixing
procedure applied at ground. For PTypes 3 and 6 it specifies which type of
samples have been differentiated on-board.

The data segment of each SCI packet is filled up to the maximum length of
980~bytes. If data are compressed, the compressor stops the processing of new
data when the maximum length is reached. The number of compressed samples is
stored in the tertiary header in order to verify the consistency of the
decompression. In case the acquisition is stopped, by sending a proper
telecommand, the buffers of the REBA are flushed by storing the remaining data
in new TM packets until the buffers are empty.  In that case, packets much
shorter than the maximum length may be generated.

At ground, packets have to be grouped according to their content and sorted by
time, an action called {\em registration} and time unscrambling. Each packet
contains all the relevant information to decode/decompress the data, register
the data source and reconstruct the appropriated OBT of the packet. A packet
contains consecutive samples produced for a given detector and generated
according to a given processing type and for a given set of processing
parameters. Any telecommand that changes one or more processing parameters for a
given detector must be preceded by a telecommand stopping the acquisition.

In order to allow proper time reconstruction of all scientific samples, the OBT
of each SCI packet corresponds to the time at which the first sample in the
packet has been acquired. The time of the other samples within the packet is
calculated based on the packet OBT and the channel frequency.

\subsubsection{Mapping HK data into packets}
The mapping of HK data into telemetry packets follows the rules provided in the
ESA PUS. No compression is applied to the data segment and each field of such
segment is either a parameter field containing a parameter value or a structured
field containing several parameter fields. The PUS only indicates the parameter
types of the parameter fields in the packet structures, such as boolean,
enumerated, signed/unsigned integers with different octet sizes, real,
character-string, absolute/relative time.

In general, in the HK packets the SID is followed by a sequence of values of
housekeeping and diagnostic parameters that are sampled once per collection
interval. The time of the samples correspond to the OBT of the packet.

In LFI, some of the HK packets are periodic, i.e. produced on a regular basis by
the instrument, while others are produced only under particular conditions. The
periodic HK packets report continuously the
parameters of the instrument. They are classified in essential HK, fast HK and
slow HK, according to their sampling rate. The essential telemetry is intended
to be delivered to ground during the operations of the instrument when the
satellite TM capability is limited by the use of the Low Gain Antenna. They are
sent once every 32 seconds. The fast telemetry is sampled once per second. To
avoid the generation of frequent short packets, some of the fast telemetry
packets are super--commutated: the data segment is split into four sub-fields
having the same structure (same parameters at the same relative offset). Each
sub-field is sampled once per second and hence the whole packet is delivered
every 4 seconds. Finally, the slow telemetry is generated once every 64 seconds.

The non-periodic telemetry includes packets which are generated only in
response to a given telecommand (non-periodic packets) or packets, such as
events and alarms, that can be automatically generated on-board when a certain
condition arises.
\FIGLEVEL1ARCH
\section{The Level 1 software system overview}\label{sec:L1:pipeline}

The LFI Level 1 software system is composed by three main subsystems (Fig.~\ref{fig:level1arch}):

\begin{itemize}
\item a {\bf Real-time Assessment} system. Based on the standard ESA SCOS-2000
  system (Spacecraft Control \& Operations System), it provides tools for
  monitoring the overall health of the instrument and detect possible anomalies
  \citep{Scos03}. The housekeeping data to be monitored include: temperature
  sensors output and stability, power consumption of individual units, cooling
  system parameters. Another set of tools is dedicated to the control of the
  instrument; they provide interfaces to select specific telecommands and set
  their parameter values, send the telecommands to the instrument control unit
  (the REBA) and receive TM reports on the execution status of each telecommand
  queued. Scripting capabilities are also provided in order to automatize and
  simplify the testing procedures during the ground test campaign.
\item a {\bf TeleMetry Handler} system.  The TMH has been mainly developed by a
  team of the ISDC data centre in Geneva, by reusing part of the software modules
  they have developed for the INTEGRAL gamma-ray mission \citep{Tur04}. It receives the raw
  telemetry directly from the instrument, during the ground tests, or from the
  Mission Operation Centre (MOC), during Flight Operations. The TMH implements
  the so called Level 1 pipeline, a sequence of processing steps that starts
  with the reception of the raw telemetry and that terminates with the
  production of SCI and HK timelines, properly calibrated and converted into
  physical units.
\item a {\bf Telemetry Quick-Look} system. The TQL system provides a set of
  graphical tools to visualize the SCI and HK data, display them in real-time
  and calculates on-the-fly quick statistical functions and fast Fourier
  transforms. The same set of tools are also used to display the archived
  data. It provides several independent views of the data by supplying
  information on the source of the SCI telemetry flow at a given instant
  (radiometer and detector) and the processing type applied to a given detector
  data, highlighting telemetry gaps and producing XY-plots for the HK and SCI
  timelines, correlation plots and histograms.
\end{itemize}
\FIGTMU
\subsection{SCOS  in the LFI LEVEL 1}\label{sec:scos2k}

\scos\ is the generic mission control system software of ESA
\footnote{www.egos.esa.int/portal/egos-web/products/MCS/SCOS2000/} and runs
under Sun/Solaris and Linux operating systems. It supports CCSDS telemetry and
telecommand packet standards, and the ESA PUS. Telemetry and telecommand packets
structure is kept in the SCOS Mission Information Base (MIB). This database is
built for each mission, by providing a set of ASCII tables (MIB tables) that
follow the format specified in the SCOS MIB Interface Control Document (ICD).

In addition to a set of configurable tasks that can be directly used to control
and monitor an instrument, \scos\ is also a reusable library of components with
an object-oriented design and using standard design patterns. ESA \scos\ is
available under an open-source licensing scheme, giving the possibility to
customize existing \scos\ tasks or develop additional components.

\scos\ is not able to process compressed scientific telemetry and the available
graphical displays have still limited functionalities. For this reason
\scos\ is mainly used as an EGSE system (during the ground test campaign)
and will also be used to monitor and visualize the LFI HK data during the flight
operations. Since SCOS provides also TCP/IP and CORBA external interfaces,
it is also used as a telemetry gateway between the instrument and the LFI TMH/TQL system.

\scos\ is in charge of receiving the telemetry directly from the TM source,
filling and maintaining a local database of raw packets, creating a list of
stored packets, checking for missing, damaged or duplicated packets, displaying
HK telemetry by converting raw data into physical data, generating and sending
telecommands to the instrument through the testing environment or the MOC,
checking OOL values and generating alarms to the operator.

The main task to configure the SCOS system for a particular instrument is the
definition of the MIB database. It involves the mapping of all the instrument HK
and telecommand packet structures into a subset of the MIB tables. For instance,
the Packets Identification table (PID) includes, for each TM packet, the APID,
Type, Subtype, and SID of that packet, representing its ``primary key''. A
monitoring Parameter Characteristics table (PCF) includes, for each HK parameter
to be monitored, the parameter ID, its description and type, a reference to the
calibration curve to be applied. A third table, the Parameter Location in Fixed
packets table (PLF), associates a parameter of the PCF table to one or more
packets of the PID table, by specifying its offset within the packet and
eventually the number of occurrences (for super-commutated packets). An
analogous procedure is followed to specify the telecommands
structure. Additional information that can be provided in the MIB are the OOL
values, that can be associated to validity criteria in order to distinguish
among different test conditions (e.g. warm or cryogenic temperatures),
conditional calibration curves, alphanumeric and graphic displays.

A dedicated software development was necessary to tailor specific
\scos\ tasks to the LFI ground segment, to allow the communication
between \scos\ and external systems. This tailoring has followed
the SCOS release updates that were necessary during the mission
preparation. For the tests involving only the LFI instrument, the
EGSE system was based on SCOS 2.3e. In such version, the Telemetry
Receiver (TMR), which is part of the SCOS 2000 simulators, has
been modified in order to replay simulated LFI HK and SCI
telemetry. To distribute the telemetry in real-time from SCOS to
the TMH/TQL system, a task using the Flight Dynamic Interface of
SCOS (FDIF) has been tailored in order to allow the TMH system to
register as a SCOS client, set the list of packets to be
distributed and receive a notification each time a live packet
arrives (Fig.~\ref{fig:level1arch}).

Starting from the Planck integration tests, the EGSE system adopted is the
HPCCS. This system uses a new application protocol, called Packet Interface
Protocol for EGSE (PIPE), to communicate with external equipments and
software. With this new version, a new software has been developed, the
LFIGatewy, acting as a telemetry gateway between the HPCCS main system operated
by the Spacecraft instrument team and the client systems of the LFI instrument
team (another HPCCS and the LFI TMH/TQL).

\subsection{The Telemetry Handler System}\label{sec:tmh}

\FIGTQL

The \TMH\ system is the core of the Level 1 pipeline \citep{Mor06}. It receives
real-time LFI telemetry from \scos\ or from the LFIGateway and
parses the content of each received TM packet to first discriminate between SCI
and HK telemetry packets. SCI packets are then grouped according to the
radiometer and detector source and to the applied processing type. For each
group, the scientific data samples are uncompressed an decoded, the OBT of each
sample is calculated based on the packet OBT and the detector frequency.  At
last, a calibration is applied to convert the raw sample value to Volts. The
output is saved as a Time Ordered Information file in FITS format. HK telemetry
packets are also grouped, according to the packet type. Their structure is
recovered from the SCOS MIB files so that each HK parameter within the packet is
properly extracted and converted into an engineering value according to its
calibration curve. Again, for each parameter the output is a TOI. HK and SCI
TOIs are then the input of the Level 2 of the LFI/DPC, where a more accurate
calibration is performed. All these processing steps are divided into three main
software components which form the pipeline.

The first component is the {\bf Data Receive} module
(Fig.~\ref{fig:level1arch}). This component receives in real-time
raw packets via a socket using the TCP/IP communication protocol.
It runs continuously by reading incoming packets and storing the
raw packets in a local TM archive, as FITS binary tables.
Concurrently, it sends the raw packets also to the TQL system.

The second component is the {\bf TeleMetry Unsrcambler} ({\bf TMU}). It runs in
near real-time and reads the raw packets from the TM archive. The TMU decodes
the HK packets, groups them according to the APID, Type, Subtype and SID values,
extracts the raw parameters values and saves them in an intermediate Unscrambled
TeleMetry archive (UTM) as FITS files. For the SCI packets, it parses the
information available in the tertiary header, groups and sorts the packets, and
stores them in the UTM archive. At this level, the scientific samples are not
yet decoded. In the UTM archive, each FITS file created covers a chunk of one
hour in order to avoid too big files.

A simplified class diagram of the TMU is shown in
Fig.~\ref{fig:tmu}. The main class, which executes the TMU event
loop, is the {\em Preprocessing} class \citep{Mor04}. It inherits
from five classes, each one implementing a subset of the
Preprocessing tasks. The {\em TaskLoop} class implements the main
{\em
  Loop}, by driving the activity of all the other components.  The {\em
  TelemetryScan} class represents the {\em Preprocessing} input; it controls the
access to the TM archive by providing a method, {\em NextPacket()}, which
returns the subsequent telemetry packet available in the FITS archive. The {\em
  Organizer} class stores the incoming packets into packet chains (represented
by the {\em PacketChain} class), using a distinct chain for each APID and
sorting the packets by their sequence counter. Data are then reorganized into
time slices named Science Windows. The {\em ScWinManager} class has the task of
detecting a new science window depending on a particular criteria
(e.g. pointing, slew, OBT interval) and the information included in specific
packets. For LFI, the science windows are based on preconfigured OBT
intervals. The {\em TelemetryParser} class processes each chain of telemetry
packets by selecting the appropriate packet parser: the {\em LFIScienceParser}
to decode SCI packets, or the {\em HkParser} to extract the parameters from the
HK packets.

The last component of the TMH is the {\bf TM2TOI} application. This component
reads the UTM archive and produces the final product which is the Time Ordered
Information(TOI) archive in FITS format.  TM2TOI gathers the data of the same
type, i.e. SCI data from the same detector and with the same processing or HK
data from the same packet type, it retrieves the sky and load data (for science)
and stores them in the TOI archive. For housekeeping data, TM2TOI produces one
FITS file for each housekeeping parameter and applies calibration curves if
necessary. TOI data are also stored by chunks of one hour and can be directly
used for data analysis.

Each TOI includes time information in OBT format and converted into
seconds. Science data are stored as sky and load samples. HK data are stored as
raw samples with the corresponding converted values. A quality flag is also
provided assuring the quality of the data.

\subsection{The Telemetry Quick-Look}\label{sec:tql}
\FIGPTYPESPROCESSING
The scope of the TQL sub-system is to perform simple analysis of
the data and display them together with the raw data in real-time.
It is a progression of the display tools used for the INTEGRAL
mission. Three displays are sufficient during the tests of the LFI
instrument to verify the performance and to control the stability
of the instrument. First the science data display, which can
display any combination of the 44 scientific output streams.
Second a mode display to verify the processing mode of each
detector and finally graphs to display the housekeeping (HK) data
in three different views: as a correlation plot of one HK
parameter against several others, as a histogram plot, and, third,
as a plot of any combination of HK parameters as a function of
time. All displays can be operated in two modes. Either in
real-time, to display the accumulated data during a ground test,
or in playback mode, to display archived data. A subset of the TQL
displays is shown in Fig.~\ref{fig:tql}.

A toolkit is part of the scientific display to perform simple
functions on the data. These functions are coded as ROOT macros
\footnote{http://root.cern.ch/drupal/}, which can be interpreted
in real-time by the program. For instance, the function displaying
the fast Fourier transform of a given scientific channel is coded
as a ROOT macro (see the FFT display in Fig.~\ref{fig:tql}): it
uses the fft function and the scientific samples exposed by the
toolkit. Additional macros calculate quick statistics (mean,
variance, skewness, kurtosis) of the displayed signal, or estimate
the gain modulation factors on-the-fly, using the received sky and
load samples. Each macro can be modified by the TQL user and new
macros can be defined and loaded at run-time with the scientific
display.

The ROOT framework is used as base library for all displays \citep{Roh04}. The
ROOT CINT interpreter, part of every program, linked with the ROOT library, is used
to interpret the toolkit macros. The capability of ROOT to store every object
in a file is used to store the configuration of the displays for later use.

\section{The Level 1 scientific data processing}

When creating the TOIs, the Level 1 pipeline has to recover most accurately the
values of the original (averaged) sky and load samples acquired on-board. The
operations of each processing step performed by the TMH are illustrated in
Fig.~\ref{fig:ptypes:processing}. Data acquired with PTypes 4, 5 and 6 first is
uncompressed. The loss-less compression applied on-board is simply inverted and
the number of samples obtained is checked with the value stored in the
tertiary header.

\subsection*{Conversion to physical units}

The digitized data, processed by the REBA, are not in physical units but in ADU
(Analog to Digital Unit). Conversion of $\overline{S}_{sky}$ and
$\overline{S}_{load}$ in Volt requires the Data Source Address
($\DSA$), indicating the radiometer and detector from which the data are
generated, the blanking time (indicized by the Blancking Time Index, $\BTI$),
the $\DGI$ and the $\DOI$. These values are recovered from the packet tertiary
header. Hence, the value in Volt is obtained as:
 %

%
 \begin{equation}
\overline{V}_{i} = \frac{\overline{S}_{i} - \DAEZero}{\DAEGain} - \DAEOffset
\end{equation}
\noindent
This conversion is the only processing required by PTypes 0 and 3 and it
is the last step in the processing of all the other processing types.

\subsection*{Dequantization and demixing}
\PTypeII\ and 5 data has to be dequantized by
 \begin{equation}
   P_i = \frac{Q_i}{\REBAq} - \REBAo;
 \end{equation}
 \noindent
and then demixed to obtain $\overline{S}_{sky}$ and $\overline{S}_{load}$\
 \begin{eqnarray}
    \overline{S}_{sky}  & = & \frac{\REBAgt \cdot \Pone - \REBAgo \cdot \Ptwo }{\REBAgt - \REBAgo}; \\
    \overline{S}_{load} & = & \frac{\Ptwo - \Pone}{\REBAgt - \REBAgo};
 \end{eqnarray}
\subsection*{Averaging}
Since \PTypeI\ data is just coadded on-board, the division by $\REBAnaver$ is
performed by the Level 1 software.

\subsection*{OBT reconstruction}

The OBT reconstruction for scientific data has to take into account the phase
switch status. If the phase switch is off, it means that the packet contains
consecutive values of either \sky\ or
\load\ samples. In this case, denoting with $\ismp \ge 0$ the sample index
within the packet, we have that:
  \begin{equation}
  t^{\mathrm{obt}}_{\ismp} = t^{\mathrm{obt}}_{0} + \ismp\frac{\Naver }{\freqsampling},
\end{equation}
 \noindent
with $t^{\mathrm{obt}}_{0}$ the time stamp of the packet and $\ismp = 0$ denoting the first sample in the packet.

\noindent
If the switching status is on, either consecutive pairs of (\sky, \load) samples
or (\load, \sky) samples are stored in the packets. Hence, consecutive couples of
samples have the same time stamp and
 \begin{equation}
  t^{\mathrm{obt}}_{\ismp} = t^{\mathrm{obt}}_{0} + 2\;\trunc\left[\frac{\ismp}{2}\right]\frac{\Naver }{\freqsampling}.
 \end{equation}

\section{Conclusions}\label{sec:res:concl}
The LFI Level 1 software systems are successfully in use since
five years to support the LFI instrument and calibration tests and
the Planck integration and validation tests. Their design,
development and testing has been performed according to the ESA
software engineering standards. Their development has been based
on already consolidated software: the ESA generic mission control
system and the INTEGRAL ground segment of the ISDC. This paper has
been focused on the software versions in use during the LFI
calibration campaign and the Planck integration tests. The
Operational Model (OM) of the Level 1 software, integrating the
interfaces with the MOC that will be used during flight operations
and a new I/O library, named Data Management Component (DMC), to
store the TOIs and additional auxiliary data into a relational
database system, has been presented in \citep{Mor08} and will be
the subject of an incoming paper.

\vspace{4mm}

\acknowledgments Planck is a project of the European Space Agency
with instruments funded by ESA member states, and with special
contributions from Denmark and NASA (USA). The Planck-LFI project
is developed by an International Consortium lead by Italy and
involving Canada, Finland, Germany, Norway, Spain, Switzerland,
UK, USA. The Italian contribution to Planck is supported by the
Italian Space Agency (ASI). The US Planck Project is supported by
the NASA Science Mission Directorate.

\bibliographystyle{JHEP}
\bibliography{dpc_level1} 

\end{document}